\documentclass[11pt]{article}
\usepackage{amssymb}
\usepackage{graphics,epsfig,subfigure}
\usepackage{diagbox}
\usepackage{amsthm}
\usepackage{amscd}
\usepackage{amstext}
\newcommand{\ba}{\begin{array}}
\newcommand{\ea}{\end{array}}

\def\be{\begin{equation}}
\def\ee{\end{equation}}
\def\ba{\begin{array}}
\def\ea{\end{array}}

\def\dalemb#1#2{{\vbox{\hrule height .#2pt
        \hbox{\vrule width.#2pt height#1pt \kern#1pt
                \vrule width.#2pt}
        \hrule height.#2pt}}}

\begin{document}

\begin{center}


\vspace{1cm} { \LARGE {\bf Chains of mini-boson stars}}

\vspace{1.1cm}

Shi-Xian Sun\footnote{sunshx20@lzu.edu.cn}, Yong-Qiang Wang\footnote{yqwang@lzu.edu.cn},
and Li Zhao \footnote{lizhao@lzu.edu.cn, corresponding author}

\vspace{0.7cm}

{\it $^{a}$Lanzhou Center for Theoretical Physics, Key Laboratory of Theoretical Physics of Gansu Province,
	School of Physical Science and Technology, Lanzhou University, Lanzhou 730000, People's Republic of China\\
	$^{b}$Institute of Theoretical Physics $\&$ Research Center of Gravitation, Lanzhou University, Lanzhou 730000, People's Republic of China}

\vspace{1.5cm}

\end{center}

\begin{abstract}
\noindent
In this paper, we re-investigate the  stationary, soliton-like solutions in the model of the
Einstein gravity coupled to  a free and complex scalar field,  which have been  known as mini-boson stars. With numerical method, we  find that in addition to the usual single mini-boson star solutions, there exist
  a novel family of solutions  interpreted
as chains of boson stars, which is  made of some boson stars along the symmetry axis. We show the configuration of  two types of chains, including  an even number of constituents
and  an odd number of constituents.
Furthermore, we also study the effect of  the frequency of the complex
scalar field on the ADM mass $M$ and the $U(1)$ scalar charge $Q$. It is interesting to note that the existence of chains of boson stars does not require the introduction of a complex scalar field with self-interacting potential.
\end{abstract}

\vspace{5cm}

\pagebreak

\section{Introduction}
It is well known that the most general stationary solution of the vacuum Einstein equation in a  four-dimensional spacetime is described by
 a so-called Kerr geometry, which is  a rotating black hole only with two parameters, the black-hole mass  and angular momentum.
 The above uniqueness of the Kerr black hole is sometimes referred to as the ``no-hair" theorem. When considering the model of
a free scalar field  minimally coupled to Einstein's gravity,
one could find that it is difficult to obtain analytical or numerical  black hole solutions for a long time.  Until the last few years,
 a novel family of solutions of Kerr black holes with  scalar hair was presented by  Herdeiro and
Radu in \cite{Herdeiro:2015gia}. Furthermore, when restrictions on the stationary solution  of compact object with  event horizon are lifted, there exist  a  family
of horizonless solutions, which have become well-known as boson stars (BSs).

The study of BSs has a long history. Firstly,  in the 1960s, a family of spherically symmetric BSs were discovered by
Kaup in the four-dimensional Einstein gravity coupled to a free, complex scalar field with the constant  angular frequency of the phase of the field in the complex plane \cite{Kaup:1968zz,Ruffini:1969qy}. Due to little astrophysical interest in directed searches for a boson star, BSs with rotational symmetry were not studied until the 1990s.
The first rotating solutions of boson stars in  the Einstein-Klein-Gordon  theory were found in \cite{Yoshida:1997qf}.
BSs   with  a  free scalar field  without self-interaction were   well known  as {mini}-boson  stars (BSs), which 
 can  then be extended to the self-interacting BSs case \cite{Kling:2017hjm}, the excited case of scalar field with nodes \cite{Collodel:2017LB}  and the multistate boson stars \cite{Bernal:2009zy,Li:2019mlk}.   There are also  studies on the solutions of the  self-gravitating solitons in Einstein-Proca or Einstein-Dirac models  \cite{Brito:2015pxa,Finster:1998ws}. Another interest of BSs in astronomy is to investigate their  application to axions \cite{Guerra:2019srj,Delgado:2020udb}, cosmic dark matter \cite{Suarez:2013iw,Hui:2016ltb}, and black hole mimickers \cite{Cardoso:2019rvt,Glampedakis:2017cgd,Herdeiro:2021lwl}.
In addition, the collisions of binary BSs have been studied extensively \cite{Palenzuela:2006wp,Palenzuela:2007dm,Helfer:2018vtq,Bezares:2017mzk,Palenzuela:2017kcg,Sanchis-Gual:2018oui,Bustillo:2020syj}, which offers a
possible way to detect the BSs with the gravitational waves  generated by  the merger of binary stars. 

Recently, by introducing  the self-interaction potential  with the type of quartic and sextic terms, which was studied in literatures on the Q-balls in the field theory, a novel family of solutions  interpreted
as chains of boson stars, which is  made of some boson stars along the symmetry axis, was obtained in  the Einstein-Klein-Gordon  theory \cite{Herdeiro:2021mol}.
Such solutions can be divided into two classes, including even chains and odd chains according to the parity at the two sides of the equatorial plane. The obvious difference is that in the case of even-numbered chains, the curve of the relationship between mass and frequency could form a spiral pattern  similar to the case of a single boson star,
 while  odd-numbered star chains can form    a  kind of loop pattern, which is very different from  the case of a single boson star. Moreover,
in the second branch of odd-numbered star chains with loop patterns, the
system of star chains turn into  a radially excited spherically symmetric boson star. 
Rotating generalizations of  chains of boson stars with the sextic potential also been studied in \cite{Gervalle:2022fze}.

It is worth noting that in flat spacetime there does not exist  the chain of  multisoliton  in the free scalar field model.
So, it will be interesting to see whether there are
solutions of the chains of  mini-boson stars without the self-interaction potential in the model of free scalar field coupled to gravity.
In the present paper, we  numerically solve the coupled system of nonlinear partial differential equations
of scalar field  and Einstein equations, and obtain a family of chains of  \textit{mini}-BSs, which can be divided into two classes, including even chains and odd chains.
Moreover, we also study the effect of  the frequency of the complex
scalar field on the  ADM  mass and the  Noether charge.

This paper is organized as follows. In Sect. \ref{sec0},
we  briefly review the model of a  free, complex scalar field coupled to Einstein's gravity. The boundary conditions are analyzed in Sect. \ref{sec1}.  We show the numerical results of two types of BSs chains in Sect. \ref{sec3}. The conclusion and discussion are given in the last section. Throughout this paper, Roman letters  a,b,c, $\ldots$ denote spacetime indices ranging from 0 to 3.

{\bf Note added:} when we are finishing this project, we notice that there
appears a paper \cite{Cunha:2022tvk}, which overlaps with our results
of chains of BSs with two constituents.

\section{The Model}\label{sec0}
Let us introduce the model of Einstein gravity coupled to a free, complex massive scalar field in four-dimensional spacetime. The Lagrangian density reads
\begin{equation}\label{action0}
\mathcal{S} =\int d^4x \sqrt{-g}\left(\frac{R}{16\pi G}-\nabla_a\psi^*\nabla^a\psi-\mu^2\psi^*\psi\right)\,,
\end{equation}
where $G$ and $\mu$  are the Newton's constant and the mass of the scalar field $\psi$, respectively.
Note that we consider {mini}-boson  stars without self-interaction potential, and  the term proportional to $\mu^2$  is known as a mass term. The above action is invariant under a global $U(1)$ transformation $\psi\rightarrow \psi e^{i \theta}$,
where $\theta$ is constant.
 Variation of the action (\ref{action0}) with respect to  the complex
 scalar field $\psi$ and the metric  could lead to the following Klein-Gordon (KG) equation of the scalar  field
\begin{equation}
\Box\psi=\mu^2\psi\,,
\label{eq:EKG2}
\end{equation}
and Einstein field equations
\begin{equation}
E_{ab}\equiv R_{ab}-\frac{1}{2}g_{ab}R=8\pi G(\nabla_{a}\psi^*\nabla_{b}\psi+\nabla_{b}\psi^*\nabla_{a}\psi- g_{ab}(\nabla_{c}\psi^*\nabla^{c}\psi+\nabla_{c}\psi\nabla^{c}\psi^*+\mu^2\psi^*\psi)).
\label{eq:EKG1}
\end{equation}
In the absence of the  complex   scalar field $\psi$, one of exact solutions of Einstein equations (\ref{eq:EKG1}),  which is  a spherically symmetric vacuum solution with the event horizon, is known as Schwarzschild  black hole.

In order to construct a family of  chains of  \textit{mini}-BSs with an axisymmetric distribution of  the  complex   scalar field $\psi$,
we adopt the following ansatz \cite{book}
\begin{eqnarray}
\label{metric-i}
ds^2=-e^{-2U}  dt^2+e^{2U}\left[e^{2k}(d\rho^2+dz^2)+P^2d\varphi^2 \right],\label{WLP coordinates}
\end{eqnarray}
where $(t,\rho,\varphi,z)$ are the cylindrical  coordinates in 3+1 dimensional spacetime, and $U$, $k$ and $P$, are unknown functions of the spatial non-angular coordinates $\rho$ and $z$ only. In order to  make use of the  numerical method in solving the chains of BSs, it is conventional to
converted cylindrical coordinates   $(\rho,z)$   into spherical coordinates  $(r,\theta)$ by the formulas
\begin{eqnarray}
\label{transf}
 r=\sqrt{\rho^2+z^2},~~\theta=\arctan \frac{\rho}{z} ~.
\end{eqnarray}
Moreover, ones  introduce the new functions  $f\equiv e^{-2U}$, $l\equiv \frac{P^2}{ r^2 \sin^2 \theta  }$, and $m \equiv e^{2k}$, 
then the  line element (\ref{WLP coordinates}) is converted into the Weyl-Lewis-Papapetrou coordinates 
\be
\label{metrans}
ds^2=-f dt^2 +\frac{m}{f}\left(dr^2+r^2 d\theta^2\right)
+\frac{l}{f}  r^2\sin^2 \theta d\varphi^2~,
\ee
where three metric functions $f$, $l$,  and $m$
are functions of the variables $r$ and $\theta$ only.

In addition,  one could introduce the following ansatz of matter fields:
\begin{eqnarray}\label{ansatz2}
\psi=\phi(r,\theta) e^{-i\omega t}.
\label{scalar_ansatz}
\end{eqnarray}
Here,   $\phi$ depend on the radial distance  $r$  and  polar angle $\theta$, and
the constant $\omega$ is the  angula frequency of the complex scalar field, which means that  scalar field possesses a harmonic time dependence.
Concerning the Einstein equations, we find  the set of five coupled partial differential equations  (PDEs) need to be solved. There are three equations for the combinations of  Einstein
equations: $E_r^r+E_\theta^\theta=0$,
$E_\varphi^\varphi=0$ and
$E_t^t=0$. The remaining two Einstein equations are considered as constraint conditions to
check the numerical accuracy of the  integral method. Taking the ansatz (\ref{metrans}) into account, we get the explicit form of  the KG equation and the combined Einstein equations
\begin{eqnarray}
\phi_{,rr}+\frac{\phi_{,\theta \theta}}{r^2}
+ \left(\frac{2}{r}+\frac{ l_{,r}}{2l} \right)\phi_{,r}
+ \left(\cot \theta+\frac{ l_{,\theta}}{2l}  \right)  \frac{\phi_{,\theta}}{r^2}
+ (\frac{\omega^2}{f}-\mu^2)\frac{m}{f}\phi=0 \ , \\
f_{,rr}+f,_{\theta\theta}{r^2}
+\frac{2f_{r}}{r}+\frac{\cot \partial_{\theta} f}{r^2}
-\frac{1}{f} \bigg(\partial_{r}f^2+\frac{f_{,\theta}^2}{r^2} \bigg)
+\frac{1}{2l} \bigg(f_{,r}l_{,r}+\frac{f_{,\theta}l_{,\theta}}{r^2} \bigg) \nonumber
\\
\qquad\qquad\qquad\qquad+16\pi G   \left(\mu^2\phi^2-\frac{2\omega^2 \phi^2}{f} \right) m=0~,
\label{eql}\\
l_{,rr}+\frac{l_{,\theta \theta}}{r^2}
+\frac{3l_{,r}}{r}+\frac{2\cot \theta l_{,\theta}}{r^2}
-\frac{1}{2l} \bigg(l_{,r}^2+\frac{l_{,\theta}^2}{r^2}\bigg)
+32\pi G   \left(\mu^2\phi^2-\frac{ \omega^2 \phi^2}{f} \right)\frac{l m}{f}=0~, 
\\
\label{eqm}
m_{,rr}+\frac{m_{,\theta \theta}}{r^2}
+\frac{m_{,r}}{r}
+\frac{m}{2f^2} \bigg(f_{,r}^2+\frac{f_{,\theta}^2}{r^2}\bigg)
-\frac{1}{m} \bigg(m_{,r}^2+\frac{m_{,\theta}^2}{r^2}\bigg) \nonumber\\
+16\pi G
      \left[ \frac{f}{m}\bigg(\phi_{,r}^2+\frac{\phi_{,\theta}^2}{r^2}\bigg) + \mu^2\phi^2-\frac{\omega^2 \phi^2}{f} \right]\frac{m^2}{f}=0~.
\end{eqnarray}
These four equations are second-order elliptic PDEs, which can be solved as a boundary value problem when appropriate boundary conditions are imposed.

\section{Boundary conditions}\label{sec1}
Before  numerically solving the  coupled PDEs, we should obtain the asymptotic behaviors of the four functions $f,l,m$ and $\phi$, which is equivalent to give the boundary conditions we need.
We will still use  the  boundary conditions by following the same steps   given in Refs. \cite{Herdeiro:2015gia,Herdeiro:2021mol}.

Considering an axial symmetry  system, we have polar angle  reflection symmetry $\theta\rightarrow\pi-\theta$ on the equatorial plane, and thus  it is convenient to consider the
 coordinate range $\theta \in [0,\pi/2] $.
 In addition,  the scalar field
can have either even or odd parity with respect to
reflections on the equatorial plane.  So,  we require
\begin{equation}
\left\{\begin{array}{c}
 \partial_\theta \phi(r, \pi/2)  = 0,   \;\;\;  \text{even parity} \\
\;\;\;\;\;\phi(r, \pi/2)  =0,\;\;\; \text{ odd  parity }
\end{array}
\right.
\end{equation}
together with the regularity condition at the origin,
\begin{equation}
\left\{\begin{array}{c}
 \partial_r \phi|_{r=0} = 0,   \;\;\;  \text{even parity} \\
\;\;\;\;\;\phi|_{r=0} =0,\;\;\; \text{ odd  parity }
\end{array}
\right.
\end{equation}
 The reflection symmetry restricts the other three functions satisfy the following boundary conditions  at $\theta=\pi/2$
\begin{eqnarray}
\partial_\theta f|_{\theta=\pi/2} =\partial_\theta l|_{\theta=\pi/2}=\partial_\theta m|_{\theta=\pi/2}= 0.
\end{eqnarray}
We give axis boundary conditions at $\theta=0$ where  regularity must be imposed   Neumann boundary conditions on
scalar field  and  the metric functions
\begin{equation}\label{abc}
\partial_\theta f(r, 0) =\partial_\theta l(r, 0)=\partial_\theta m(r, 0)=\partial_\theta\phi(r, 0)=0.
\end{equation}
In addition, the asymptotic behaviors near the boundary $r\rightarrow\infty$ are
\be
\label{bcinf}
f\bigl.\bigr|_{r\to \infty}=m\bigl.\bigr|_{r\to \infty}
=l\bigl.\bigr|_{r\to \infty}=1, \;\;\; \phi\bigl.\bigr|_{r\to \infty}=0
\ee
where the  Minkowski spacetime background is approached.

The chains of mini-BSs possess two global
``charges": the ADM mass $M$ and the Noether charge $Q$.  The ADM mass $M$  can be obtained in two ways: (i) from the asymptotic behavior of the solutions.
Near the boundary  $r\rightarrow\infty$, the metric function  $f$   has the following forms
\begin{eqnarray}
\label{asym}
f \rightarrow 1-\frac{2GM}{r}+\cdots,
\end{eqnarray}
where the parameters $M$ is the ADM  mass  of the BSs; (ii) from the respective Komar expressions \cite{wald},
\begin{equation}
\label{komar}
{M} = \frac{1}{{4\pi G}} \int_{\Sigma}
 R_{\mu\nu}n^\mu\xi^\nu dV~.
\end{equation}
where $\Sigma$ denotes a  spacelike hypersurface
(with  the  volume element $dV$), $n^{\mu}$ is unit vector normal to $\Sigma$ and  $\xi^\nu$ is
the everywhere timelike Killing vector field.
Besides the ADM  mass, the Noether charge $Q$  is given by
\begin{equation}
Q =  \int_{\Sigma}
 j_{\mu}n^\mu dV~,
\end{equation}
where  $j_{\mu}$ is  Noether 4-current  associated
with $U(1)$ symmetry of the complex scalar field,
\begin{equation}
j_{\mu}= -i(\psi\partial_{\mu}\psi^{*}-\psi^{*}\partial_{\mu}\psi).
\end{equation}
\begin{figure}[t]
	\begin{center}
		\includegraphics[width=0.325\textwidth]{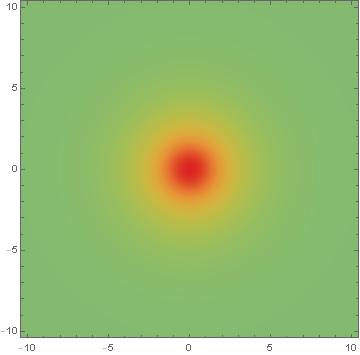}
		\includegraphics[width=0.325\textwidth]{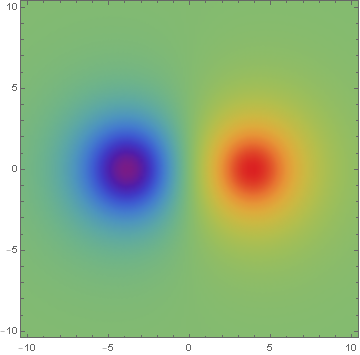}
		\includegraphics[width=0.325\textwidth]{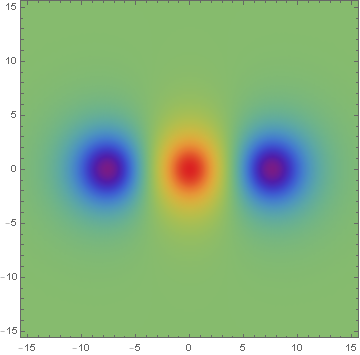}
		\includegraphics[width=0.325\textwidth]{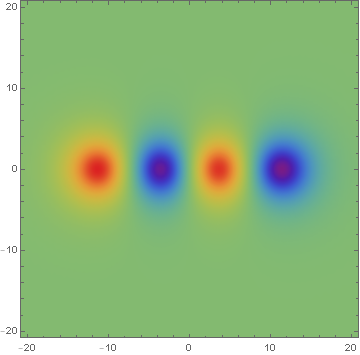}
		\includegraphics[width=0.325\textwidth]{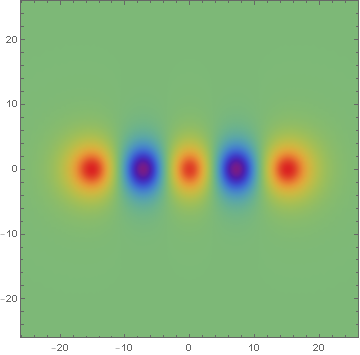}
	\end{center}
	\caption{Plots of the scalar field amplitude $\phi$ in  coordinates $\rho=r\sin\theta$ and $z=r\cos\theta$, for chains of BSs with one to five constituents.  The horizontal direction is the $z$-axis, while the vertical direction is the $\rho$-axis. The   bright spots with red and blue colour  indicate the positive and negative   value for the  amplitude $\phi$ of scalar field, respectively. The parameter is set to $\omega=0.9$.}
	\label{config1}
\end{figure}

\begin{figure}[t]
	\begin{center}
		\includegraphics[width=0.325\textwidth]{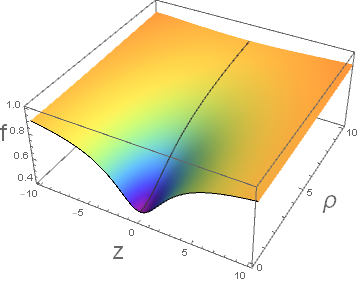}
		\includegraphics[width=0.325\textwidth]{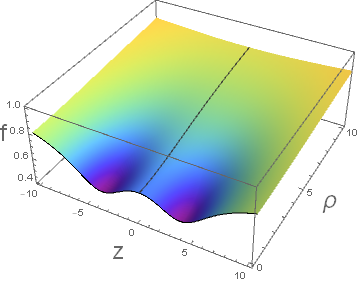}
		\includegraphics[width=0.325\textwidth]{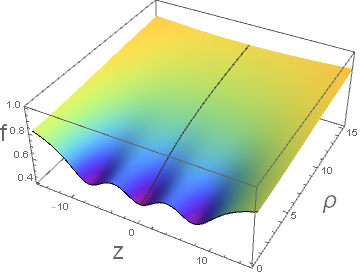}
		\includegraphics[width=0.325\textwidth]{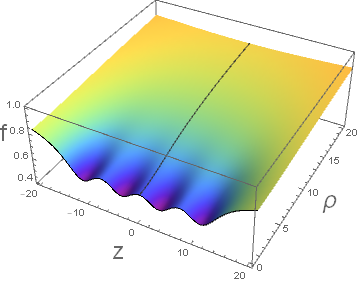}
		\includegraphics[width=0.325\textwidth]{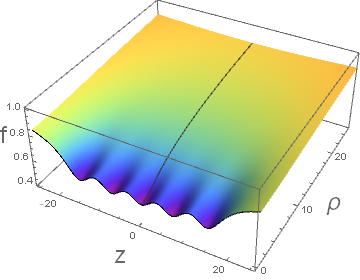}
	\end{center}
	\caption{Plots of the metric functions $f$  in   coordinates $\rho=r\sin\theta$ and $z=r\cos\theta$, for chains of BSs with one to five constituents. The parameter is set to $\omega=0.9$.}
	\label{config2}
\end{figure}
\section{Numerical results}\label{sec3}

In this section, we will solve the above  second-order elliptic PDEs  numerically. We introduce a new compactified
radial  coordinate $x=\frac{r}{r+1}$,
which maps the semi-infinite interval $r\in[0,\infty)$ to a finite
one $x\in[0,1]$. Considering the reflection symmetry $\theta\rightarrow\pi-\theta$, it suffices to consider the range of angular variables $\theta\in[0,\frac{\pi}{2}]$.
 All numerical calculations are based on the finite element method, performed by using the professional PDE solver  Comsol. Typical grids range between $80\times100$ and $100\times100$ in the integration region $0 \leq x \leq 1$ and $\theta\in[0,\frac{\pi}{2}]$. Our iterative process is
the Newton-Raphson method, and the relative error for the numerical solutions in this work is
estimated to be below $10^{-5}$.

In  the numerical results below,  we have fixed the values $\mu=1$ and $G=1/8\pi$. The only remaining
input parameter is the scalar field's frequency $\omega$. The parity of the scalar field is related to the spatial distribution of the energy density
under reflections along the equatorial plane. Due to the symmetric or antisymmetric  property under reflection, the value of parity corresponding to even number of constituents is negative, while the value of parity corresponding to  odd number of constituents  is positive.

 For both the  even-parity and odd-parity configurations, we plot the distribution of  scalar field   function on the {first} branch in Fig. \ref{config1}, which indicates chains of mini-boson stars with one to five constituents.  The   bright spots with red and blue colour  indicate the positive and negative   value for the  amplitude $\phi$ of scalar field, respectively.  
From the first panel of
Fig.  \ref{config1}, it is obvious that single {mini}-BSs  are uniformly distributed. With the increasing of the constituents of chains of boson stars, 
the deviation of the shape of each constituent  from spherical symmetry  become  grow.
 As a comparison, in Fig. \ref{config2}, we give the 3-dimensional  plots for the metric function $f(z,\rho)$, where the configuration shows alternating peaks and troughs all located symmetrically along the $z$ axis. We could see that
because of the   the distribution of the constituents of chains of boson stars , the metric functions  $g_{tt}$ is influenced at the corresponding position.

\subsection{Even constituents}
\begin{figure}
\begin{center}
	\includegraphics[height=.245\textheight,width=.33\textheight, angle =0]{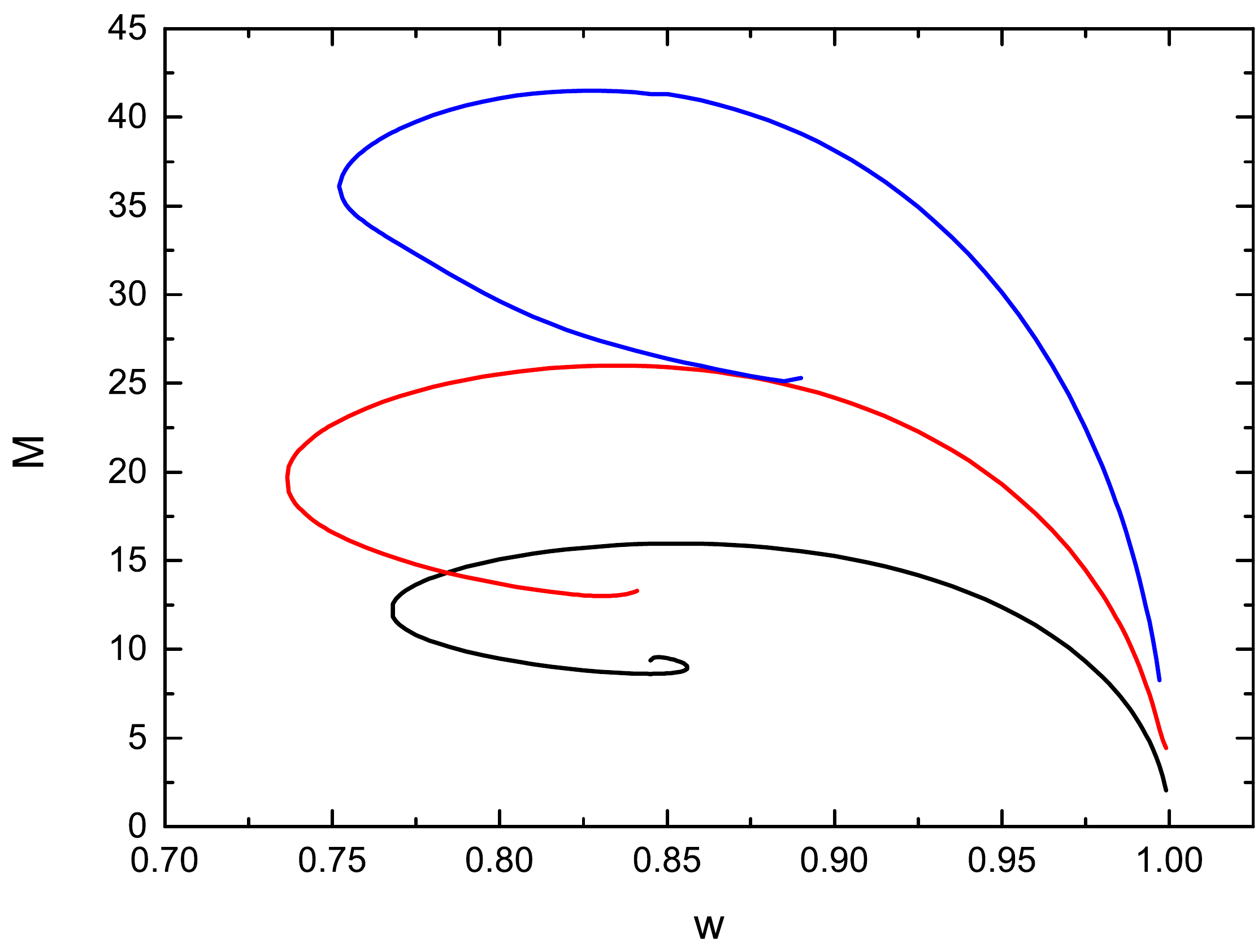}
	\includegraphics[height=.245\textheight,width=.33\textheight, angle =0]{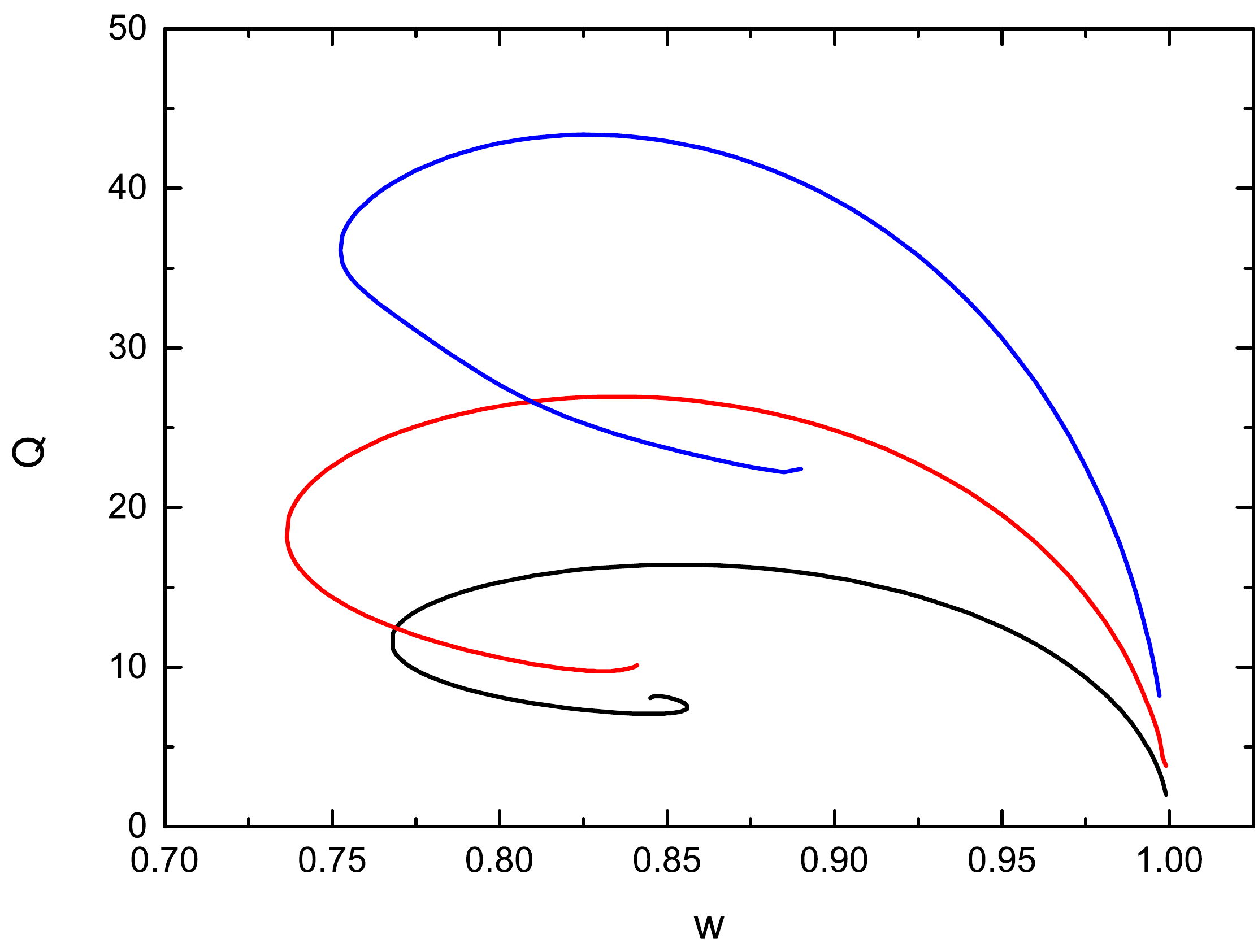}
	\includegraphics[height=.245\textheight,width=.33\textheight, angle =0]{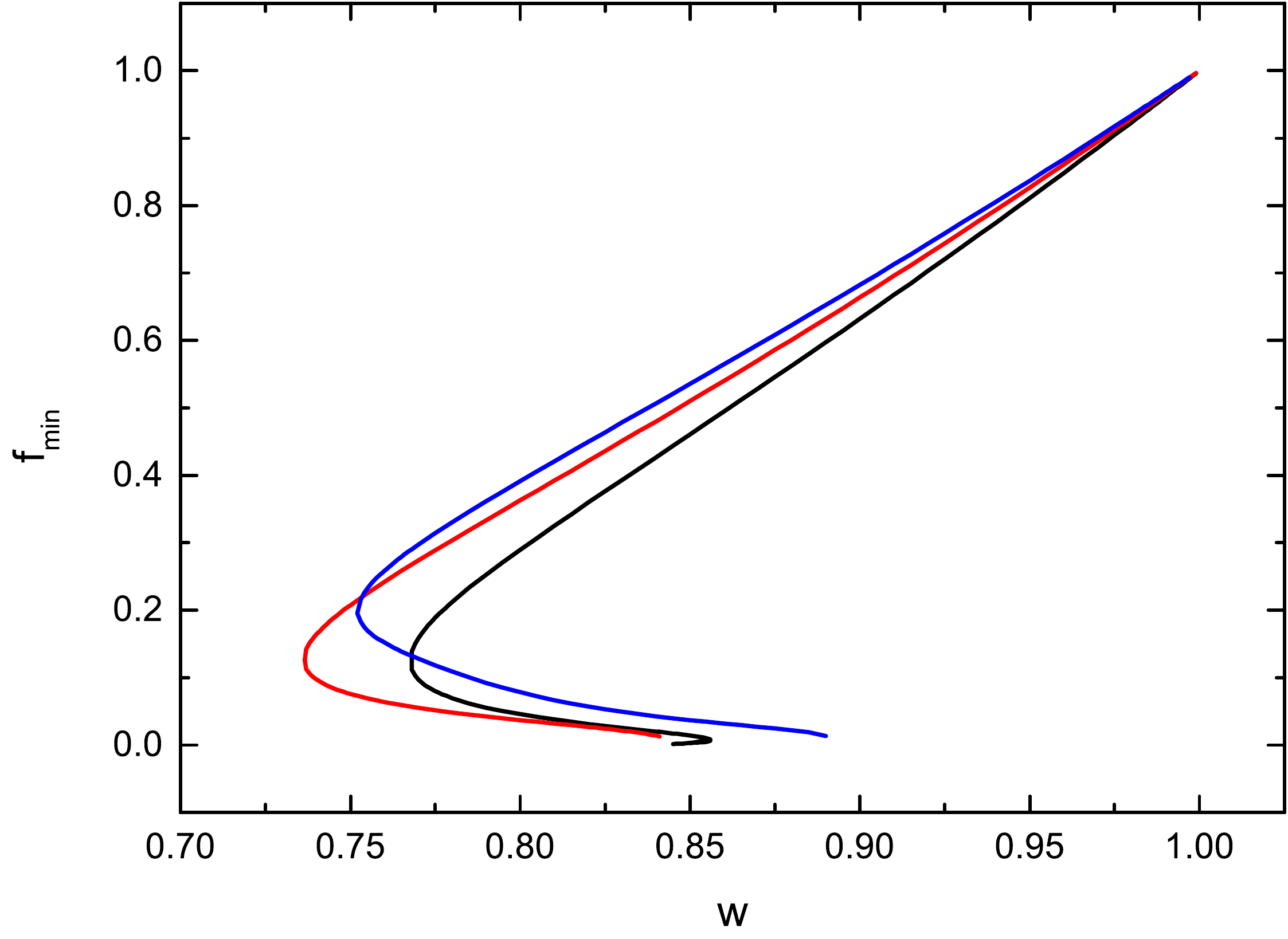}
	\includegraphics[height=.245\textheight,width=.33\textheight, angle =0]{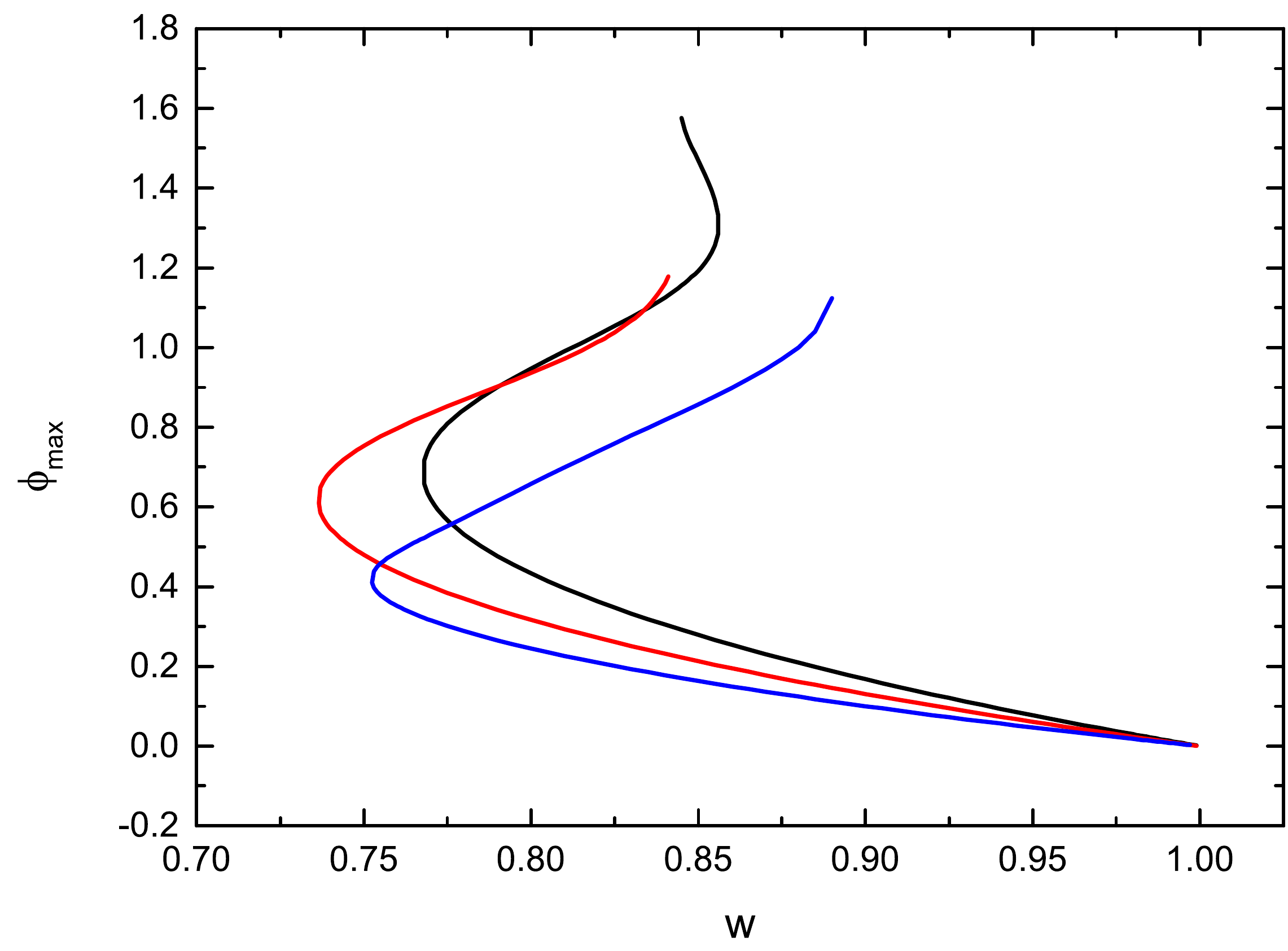}
\end{center}
	\caption{ Plots of  the dependence of ADM mass $M$, Noether charge $Q$, minimal value of the metric function $f_{min}$ and maximal value of the scalar field $\phi_{max}$ on the frequency $\omega$.  In all plots, the black,  red  and  blue lines denote  one, two  or four constitutes of BSs, respectively.}
	\label{kstability1}
\end{figure}

In this subsection, we will analyze the solutions of  BSs with  even number of constituents located symmetrically along the symmetry $z$-axis. These configurations represent chains of  BSs  have an odd-parity scalar field. The $\omega$-dependences for
even chains are shown in Fig. \ref{kstability1}. In all panels, the red  and blue lines indicate chains of BSs with two constituents and with  four constituents. 
In order to compare with the results of the single boson stars,
we also drew a   black line in each panel, which represents  the $\omega$-dependences for the single boson stars.

In the upper left panel of  Fig. \ref{kstability1} ,
we exhibit the mass $M$ versus the frequency $\omega$ for three sets of chains of BSs with one, two and four constituents, respectively. Meanwhile,
The Noether charge $Q$ as a function of the frequency  $\omega$ is shown in the right panel.
We can see  that there exist chains of BSs with even constituents 
for $\omega < \mu$, which is similar to the case of the
single boson stars.
 It is shown that, with the decrease of $\omega$, the mass of chains of BSs with even constituents  increases firstly and  reaches a maximum value of mass, and then decreases to a  minimal value of  $\omega$. Further increasing the value of $\omega$,   one can find another branch of chains of BSs.
As the frequency $\omega$ changes, the masses and charges of two or four constituents exhibit spiraling behavior similar to the single BSs.  Three spiral curves  start from the vacuum
 and revolve into the 
 center of the graph. Besides, when these curves spiral into the central region, the error in the numerical computations begins to increase, thus a finer mesh should be needed. 
The lower two panels show  the maximal  values of the scalar field function $\varphi_{max}$ and the minimal metric function $f_{min}$  at the center of the BSs,  diverges or vanishes respectively  when the frequency is increasing large. As illustrated in Fig. \ref{kstability1}, the typical feature of the higher even  number of constituents shows a similar pattern as the single BSs.

\subsection{Odd constituents}
\begin{figure}
\begin{center}
	\includegraphics[height=.365\textheight,width=.5\textheight, angle =0]{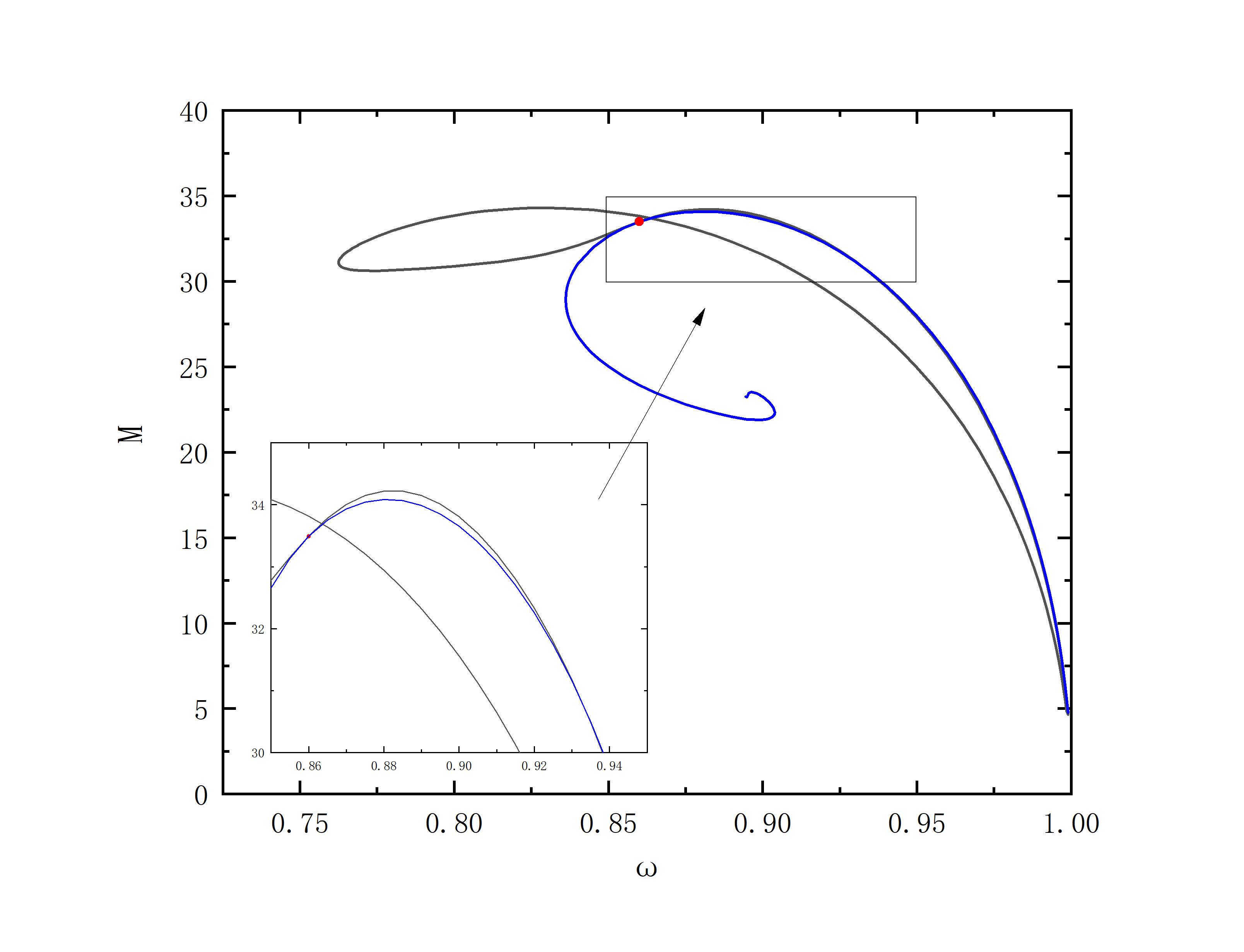}
	\includegraphics[height=.245\textheight,width=.33\textheight, angle =0]{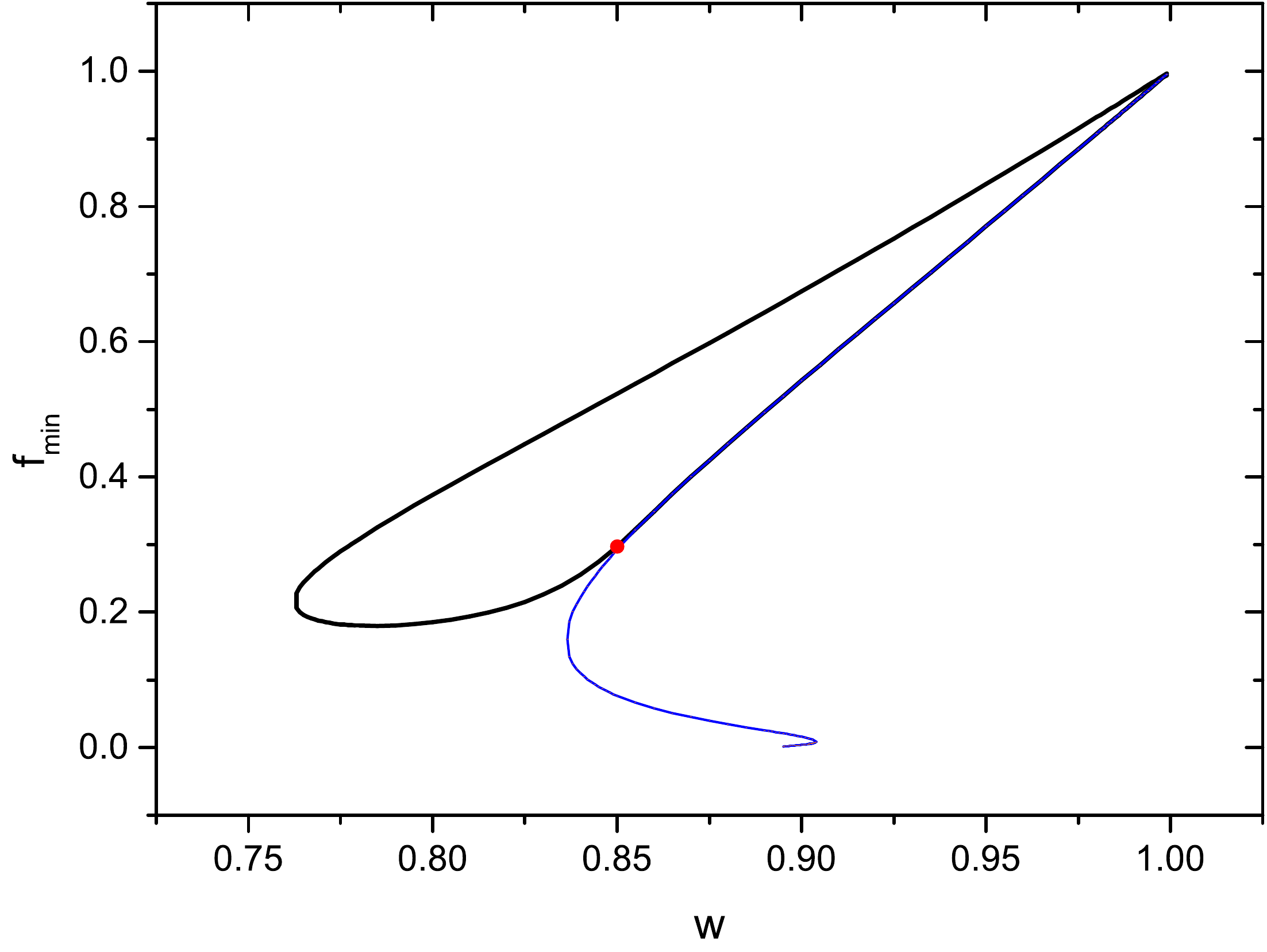}
	\includegraphics[height=.245\textheight,width=.33\textheight, angle =0]{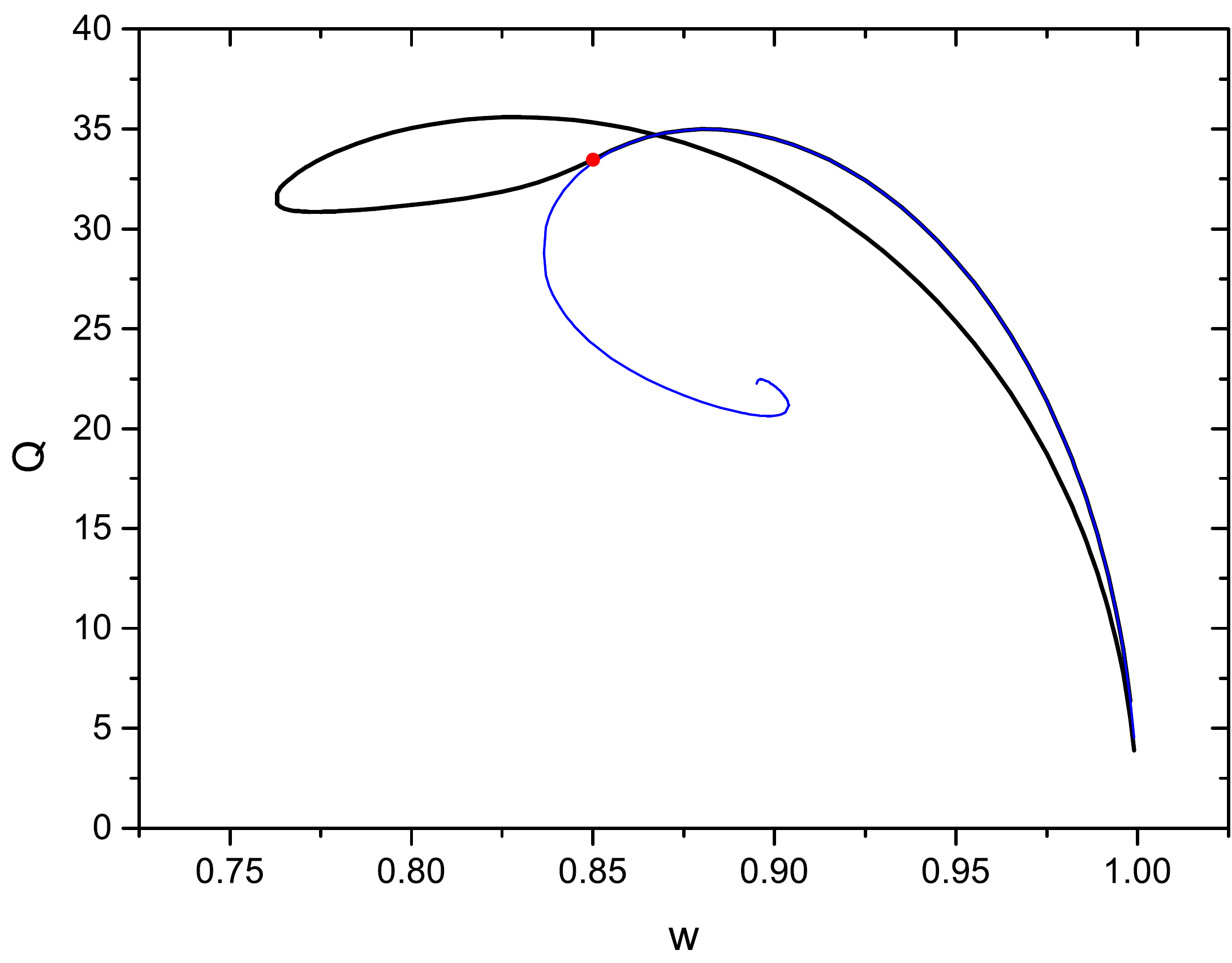}
	\includegraphics[height=.245\textheight,width=.33\textheight, angle =0]{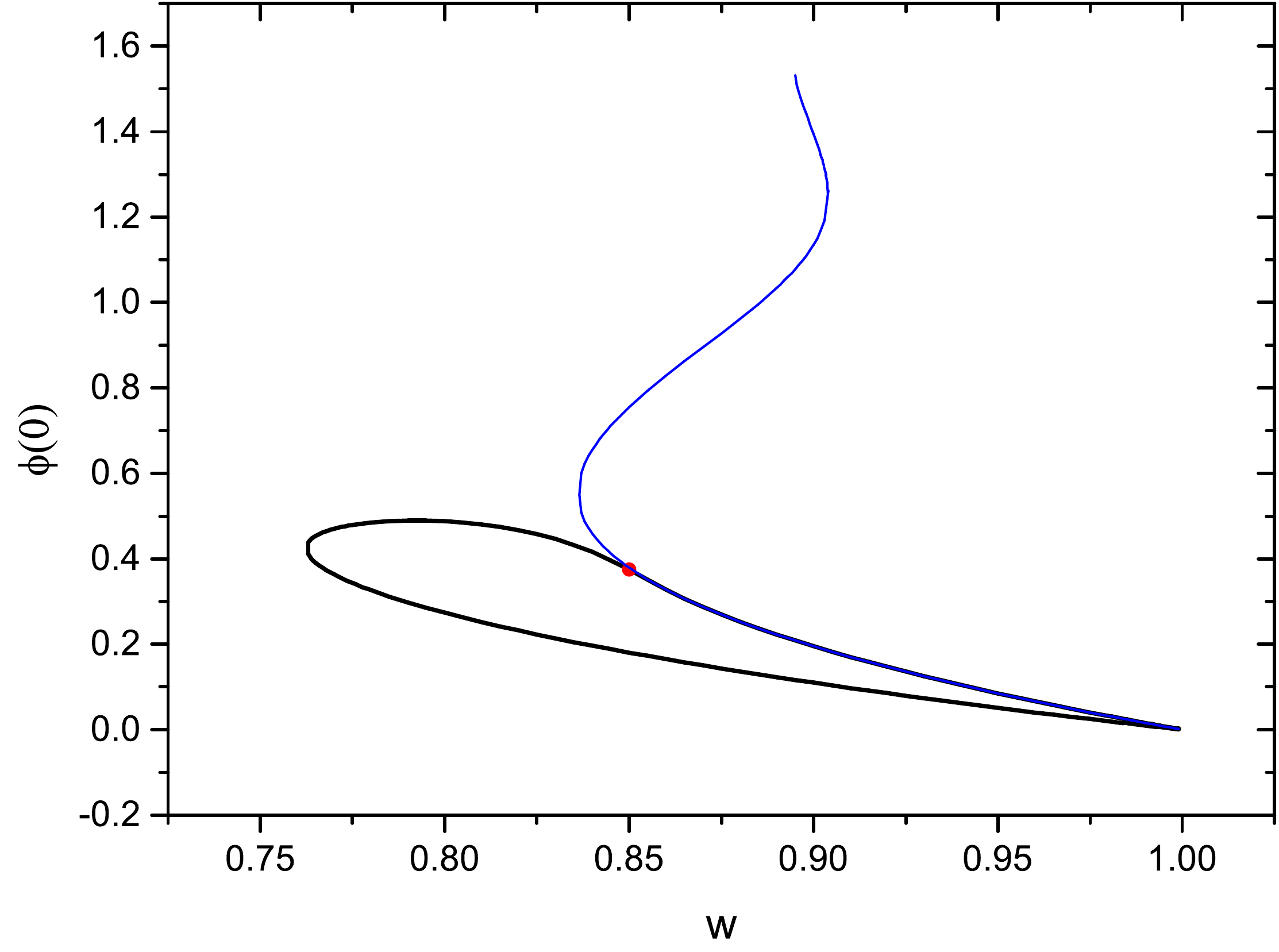}
\end{center}
	\caption{Plots of  the influence of frequency on ADM mass $M$, minimal value of the metric function $f_{min}$, Noether charge $Q$, maximal value of the scalar field $\phi(0)$. The inset in the upper left panel shows an enlarged image of one  area.
 In all plots, the black and  blue lines denote chains of BSs and first excited state, respectively.}
	\label{kstability2}
\end{figure}

In this subsection, we turn to the solutions of odd number of constituents with an even-parity scalar field. In general, the scalar field solution of $n$-th excited state possesses $n$ nodes along the radius coordinate. The profile of homogeneous scalar field solution of the first excited
state is given by an initial guess that has one node along the radius coordinate.
 In Fig. \ref{kstability2}, we present the frequency dependence of chains of BSs with three constituents, which is draw with black line. In order to compare with the results of the single boson stars,
 we also drew a   blue line in each panel, which represents  the $\omega$-dependences for  a radially excited spherically symmetric BS with a single radial  node.
  Subfigures on upper, middle and lower correspond to  the influence of frequency on ADM mass $M$, Noether charge $Q$, minimal value of the metric function $f_{min}$ and maximal value of the scalar field $\phi_{max}$.  Interestingly, the properties of such chains are
very different from those of single BS,  the $(\omega, Q)$ and $(M, Q)$-diagrams for chains of BSs with three constituents  have a loop structure. In this
case,  the loop is formed by extending from the second branch to the first branch at the upper limit of frequency critical value. 

\begin{figure}[t]
\begin{center}
\includegraphics[width=0.325\textwidth]{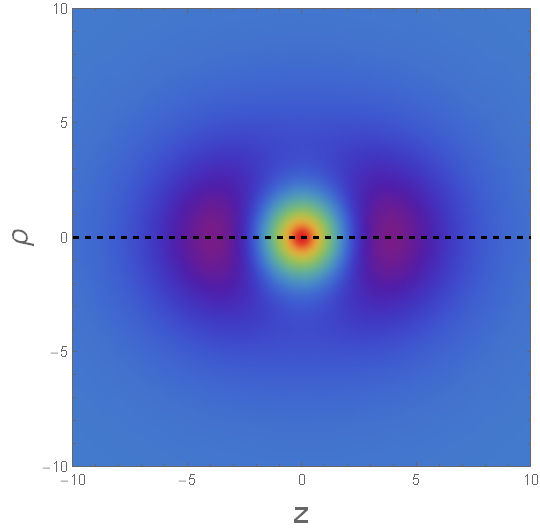}
\includegraphics[width=0.325\textwidth]{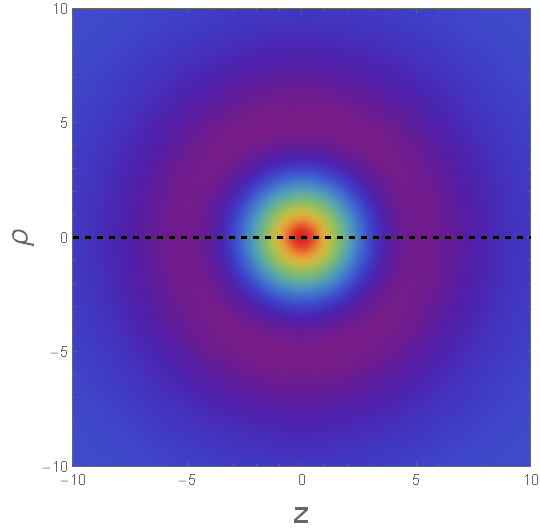}
\includegraphics[width=0.325\textwidth]{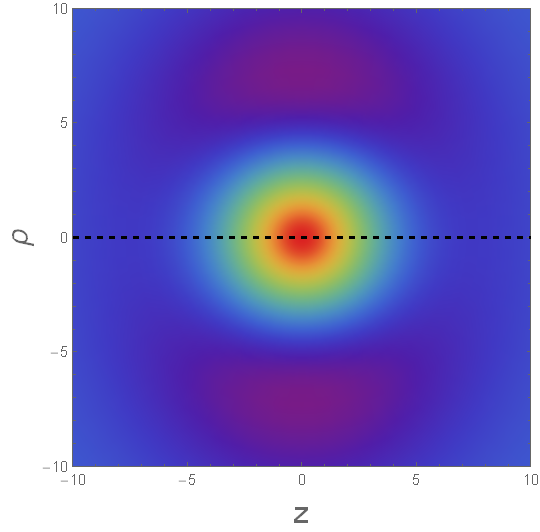}
\end{center}
\caption{Plots of the distribution of  scalar field $\phi$ with three constituents  on the second branch in   coordinates $\rho=r\sin\theta$ and $z=r\cos\theta$.  The dotted lines show the $z$ axis of symmetry. The parameter is set to $\omega=0.83$, $0.863$, $0.89$, respectively.}
\label{config3}
\end{figure}

From the  panels in  Fig. \ref{kstability2}, we see that the chains overlap on their \textit{second} branch
with a first excited spherical single BS. The red dot is the only point where they intersect, and the corresponding value of $\omega$ is 0.863. More details are shown in the inset, one could see that the two lines are close but do not coincide.
When $\omega$ deviates a little from 0.863,  the images of the scalar field  distribution  are   plotted  in the left and right subfigures  shown in Fig. \ref{config3}. From the figure, we can clearly see  the result of this model, when $\omega$ is below 0.863, the low-frequency  distribution of chains of BSs is squeezed in the axial direction; after that,  in the case that $\omega$ equals to 0.863, the central dominant of chains of BSs will be surrounded by one ``boson shell'', which means the system tends to be spherically symmetric.   When  $\omega$ is greater than 0.863, the distribution of chains of BSs tends to be on the  direction of  the equatorial plane.

\section{Discussion}
In this paper, we investigated  the  chains of BSs within a model with a free complex scalar field coupled to gravity, and obtain the bound  state of chains of {mini}-BSs with up to five constituents located symmetrically along the symmetry axis.   It is interesting to note that the existence of chains of boson stars does not require the introduction of a complex scalar field with self-interacting potential.

We have numerically solved the partial differential equation for chains of \textit{mini}-BSs by using the finite element method. It is shown that the number of extreme values of the scalar field determines the number of components in the BSs chains.
 In similar to a single ground state BS, the curves of the ($M$, $\omega$) and ($Q$, $\omega$) of the chains with an even number of constituents show a spiraling behavior. The character of the scalar field is that it is just located at the location of the sharp troughs of the metric function $f$.
On the other hand, for the odd number of constituents, we observe the curves between ADM mass and  Noether charge  in terms of frequency form the loop structure, which is different from the
case of a single spherical BSs. On the second branch, there is only one intersection with the spiraling curve of the first excited state boson star,  where the central  BS is  surrounded by one ``Boson shell".  

There are some interesting extensions of this work. Firstly, the generalization of stationary solution of chains of rotating BSs are also analyzed in Ref. \cite{Gervalle:2022fze}, 
we would like to extend our study
 to construct the rotating generalizations of  chains of boson stars with the free scalar field.
  Secondly, it is interesting 
  to construct the multistated chains of BSs, where  coexisting states of
 the several scalar fields are presented, including the ground and excited states.  Finally,
 the binding energy and stability of chains of BSs is still not discussed in this paper, which will be leaved to our further paper.

\subsection*{Acknowledgements}
This work is supported by National Key Research and Development Program of China (Grant No. 2020YFC2201503) and  the National Natural Science Foundation of China (Grants No.~12275110 and No.~12047501). Parts of computations were performed on the shared memory system at institute of computational physics and complex systems in Lanzhou university.

\end{document}